\documentclass[12pt]{article}\usepackage{amsmath,amsfonts, epsfig, lscape} \begin{document}

\def\const{\hbox{\rm const}} \def\diag{\hbox{\rm diag}} \def\mod{\hbox{\rm mod}} \def\rank{\mathop{\hbox{\rm rank}}}
\def\Image{\mathop{\hbox{\rm Image}}} \def\mes{\mathop{\hbox{\rm mes}}} \def\grad{\mathop{\hbox{\rm grad}}}
\def\ind{\mathop{\hbox{\rm ind}}} \def\exp{\mathop{\hbox{\rm exp}}} \def\d{\mathop{\hbox{\rm d}}} \def\k {\mathop{\hbox{\rm
k}}} \def\ad{\hbox{ad}} \def\Ker{\mathop{\hbox{\rm Ker}}} \def\Im{\mathop{\hbox{\rm Im}}} \def\id{\mathop{\hbox{\rm id}}}
\def\Tr{\mathop{\hbox{\rm Tr}}} \def\so{\mathop{\hbox{\rm so}}} \def\Re{\mathop{\hbox{\rm Re}}} \def\Im{\mathop{\hbox{\rm
Im}}} \def\scirc{\mathbin{\hbox{\scriptsize$\circ$}}} \def\M{\mathop{${\mathbb R}^2_1$}} \def\sgn{\mathop{\hbox{\rm sgn}}}
\def\arctanh{\mathop{\hbox{\rm arctanh}}} \def\bX{\mathop{\hbox{\bf X}}} \def\bV{\mathop{\hbox{\bf V}}}
\def\bK{\mathop{\hbox{\bf K}}}

\newtheorem{con}{Conjecture}[section]
\newtheorem{prob}{Problem}[section]
 \newtheorem{prop}{Proposition}[section]
\newtheorem{coro}{Corollary}[section]
 \newtheorem{lem}{Lemma}[section]
 \newtheorem{theo}{Theorem}[section]
\newtheorem{defi}{Definition}[section]
 \renewcommand{\theequation}{\thesection.\arabic{equation}}
  \newtheorem{cri}{Sufficient condition} \renewcommand{\thecri}{\arabic{cri}.}  \newenvironment{criterion}{\begin{cri}\rm}{\end{cri}}
\newtheorem{rem}{Remark}[section] \newtheorem{exa}{Example}[section] \newtheorem{hyp}{Hypothesis}[section]
\newenvironment{hypothesis}{\begin{hyp}\rm}{\end{hyp}} \renewcommand{\therem}{\thesection.\arabic{rem}}
\newenvironment{remark}{\begin{rem}\rm}{\end{rem}} \newfont{\gothic}{eufm10 at 12pt}

 \title{An invariant classification of cubic  integrals of motion}

\author{Raymond G.  McLenaghan \\ Department of Applied Mathematics \\ University of Waterloo, Waterloo,
Ontario\\ Canada, N2L 3G1,\\ Roman G.  Smirnov\footnote{Current address:
Department of Applied Mathematics, University of Waterloo, Waterloo, Ontario,
Canada, N2L 3G1} \\ Universit\"at Paderborn, Fachbereich 17 Mathematik
\\ Pohlweg 98, D-33098 Paderborn \\ Germany \\ and\\ Dennis The \\ Department
of Mathematics, University of British Columbia\\ 121-1984 Mathematics Rd.,
Vancouver, British Columbia\\ Canada V6T 1Z2}

\date{ } \maketitle

 \begin{abstract}
  We employ an isometry group invariants approach
 to study  Killing
  tensors of valence three defined in the Euclidean plane.
  The corresponding invariants are found to be homogeneous
   polynomials of the parameters of the vector space of the Killing
   tensors. The invariants are used to classify the non-trivial
   first integrals of motion which are cubic in the momenta of
   Hamiltonian systems defined in the Euclidean plane.
    The integrable cases isolated by Holt and Fokas-Lagerstr\"{o}m are investigated from
this viewpoint.
 \end{abstract}

\section{Introduction}
 The study of Hamiltonian systems
 with two degrees of freedom admitting first integrals of motion which are cubic in the momenta has been for many years an active
area of research \cite{Dr, FL, Ho, NY}. It is an integral part of the
well-established theory of superintegrable systems which originated in the
pioneering works by Winternitz {\it et al} \cite{FMSUW, FMSUW1} (see, for example,
\cite{Ra,Ts, WG}).

The main problem for Hamiltonian systems with two degrees of freedom can
be characterized as follows.  Let $(M, {\bf g})$ be a two-dimensional
pseudo-Riemannian manifold.
Consider a Hamiltonian system on $M$ defined by a natural Hamiltonian of
the form \begin{equation} \label{H} H({\bf q}, {\bf p}) =
\frac{1}{2}g^{ij}({\bf q})p_ip_j + V({\bf q}), \end{equation}
 via the
canonical Poisson bi-vector ${\bf P}_0 = \frac{\partial}{\partial
q^i}\wedge \frac{\partial}{\partial p_i},$ $i=1,2$, where $g^{ij}$ are the
components of the metric tensor $\bf g$ and $({\bf q},{\bf p}) \in T^*M$
are the standard position-momenta coordinates.  We wish to describe the
Hamiltonian system defined by (\ref{H}), or, more specifically, by its
potential function $V$, whose complete integrability is assured by the
existence of an additional first integral of the form:
 \begin{equation}
F({\bf q}, {\bf p}) = L^{ijk}({\bf q})p_ip_jp_k + B^{k}({\bf q})p_{k},
\quad i,j,k = 1, 2, \label{F} \end{equation}
where $L^{ijk}$ denote the
components of a Killing tensor ${\bf L}$ of valence three defined on $(M,
{\bf g})$ (see below).  The problem appears to be very difficult, since
there are three quantities involved, namely the Killing tensor $\bf L$,
vector field $\bf B$ and potential $V$ related via the partial differential
equations stemming from
the vanishing of the Poisson bracket $\{H,F\} = {\bf
P}_0\mbox{d}H\mbox{d}F = 0,$  all of which may be varied.  Most of the
known Hamiltonian systems with the designated property have been found
by fixing the form of the potential $V$ in (\ref{H}) and then trying to
solve the system of equations determined by $\{H,F\} = 0$ under this condition
(see, for instance, \cite{FL,NY}).

In this paper we consider the problem from a different perspective, which
is based on the classification of  all admissible, non-trivial
 Killing tensors defining the leading term in (\ref{F}).
This can be done by making use of a new method developed by the authors
in \cite{MST, MST1, MST2, MST3} which is based on group invariants of Killing
tensors defined on pseudo-Riemannian spaces of constant curvature under
the action of the isometry group.  The group of isometries of $M$ is the
most natural choice for such a classification, since the action of the
group preserves the dynamics of the system defined by (\ref{H})-(\ref{F}).
In this paper we are mainly concerned with the case $M = {\mathbb R}^2$,
where ${\mathbb R}^2$ denotes the Euclidean plane.  Accordingly,  we find the isometry group invariants of Killing tensors of
valence three defined in ${\mathbb R}^2$ and split the space of Killing
tensors into invariant subspaces whose respective Killing tensors are
characterized by the invariants of the isometry group. This procedure
entails a classification of  first integrals of motion which are cubic in the momenta
of Hamiltonian systems defined by (\ref{H}) in ${\mathbb R}^2.$

\section{Isometry group invariants of vector spaces of Killing tensors}

Recall that the classical theory of algebraic invariants  emerged in the 19th
century as the study of invariant properties of algebraic polynomials (see
\cite{Hil, Ol} for more details). Loosely speaking, its main problem is to
find invariants and covariants of vector spaces of homogeneous polynomials
 under a change of variables induced by a group action.

The authors have incorporated these classical ideas into the study of Killing tensors defined in
pseudo-Riemannian spaces of constant curvature \cite{MST, MST1, MST2, MST3}.
More specifically, let $(M, {\bf g})$ be an $n$-dimensional pseudo-Riemannian
manifold of constant curvature. A vector space ${\cal K}^p(M)$ of Killing
tensors of valence $p \ge 1$ defined on $(M, {\bf g})$ can be
considered as a natural counterpart of a vector space of homegeneous
polynomials of fixed degree in the classical invariant theory.  Indeed,
for a fixed $p \ge 1$ the dimension $d$ of  ${\cal K}^p(M)$ is determined
by the {\it Delong-Takeuchi-Thompson formula} \cite{De, Ts, Th}:
\begin{equation}
d= \mbox{dim}{\cal K}^p(M)=\frac{1}{n}{n+p \choose p+1}{n+p-1 \choose p}, \quad p \ge 1,
\label{DTT}
\end{equation}
where $\mbox{dim}M = n$. From this point of view, we treat each element ${\bf K} \in {\cal K}^p(M)$
as an algebraic object determined by its $d$ parameters $a_1, \ldots, a_d$.
This approach to the study of Killing tensors differs from the more
conventional approach according to which a Killing tensor is viewed as a sum of
symmetrized tensor products of Killing vectors. Our primary focus has been on considering
the behaviour of ${\cal K}^p(M)$ under the induced action of the isometry group, which
reveals invariant properties of the former. Thus, it has been possible to
describe the space of invariants, or, polynomial functions of the parameters
$a_1, \ldots, a_6$ that remain fixed under the action of the isometry group
for the vector spaces ${\cal K}^2({\mathbb R}^2)$ and ${\cal K}^2({\mathbb R}^2_1)$
(see \cite{MST} and \cite{MST3} respectively).
The new invariants turn out to be very useful in classification problems
involving Killing tensors. In this paper we turn our attention to
the vector space ${\cal K}^3({\mathbb R}^2)$ and the corresponding invariants of
the isometry group to be used  in classification of cubic integrals of motion.

\section{Cubic integrals of motion}

Consider a Hamiltonian system defined by (\ref{H}).  Assume it admits an
additional first integral of the form
\begin{equation} G({\bf q}, {\bf p})
= L^{ijk}({\bf q})p_ip_jp_k + K^{ij}p_ip_j({\bf q}) + B^{\ell}({\bf
q})p_{\ell} + U({\bf q}), \label{G} \end{equation}
where $i,j, k = 1,2.$
The vanishing of the Poisson bracket $\{H, G\} = 0$ decouples into the
following conditions expressed in component (contravariant and covariant)
and coordinate-free forms
respectively:
 \begin{eqnarray}
 L^{(ijk}_{,f}g^{\ell)f} -
\frac{3}{2}L^{f(ij}g^{k\ell)}_{, f} = 0  & \Leftrightarrow & [{\bf L},{\bf
g}] = 0, \label{1}\\ [0.3cm]
K^{(ij}{,f}g^{k)f} - K^{f(i}g^{jk)}_{,f} = 0  &
\Leftrightarrow & [{\bf K},{\bf g}] = 0, \label{2}\\ [0.3cm]
B^{(i}_{,f}g^{j)f} -
\frac{1}{2}B^fg^{ij}_{,f} - 3L^{fij}V_{,f} = 0  & \Leftrightarrow & [{\bf
B},{\bf g}] = 3{\bf L}\mbox{d}V, \label{3}\\ [0.3cm]
 U_{,f}g^{fi} - 2K^{fi}V_{,f}
= 0 \, (U_{,i} = 2K^j{}_iV_{,j}) & \Leftrightarrow & \mbox{d}U{\bf g} = 2{\bf K}\mbox{d}V, \label{4}\\ [0.3cm]
B^fV_{,f} = 0 & \Leftrightarrow & {\bf B}(V) = 0, \label{5}
 \end{eqnarray}
 where ${}_{,f}$, $($ , $)$ and $[$ , $]$ denote partial differentiation,
symmetrization and the Schouten bracket respectively.  Note $[{\bf B},
{\bf g}] = -{\mathcal L}_{\bf B}{\bf g},$ where ${\mathcal L}$ denotes the Lie
derivative operator.  It follows immediately from (\ref{1}) and (\ref{2})
that $\bf L$ and $\bf K$ are Killing tensors of valence three and two
respectively, while $\bf B$ - in general -  is not.  Equation (\ref{5}) reveals that the potential
function $V$ is preserved by the vector field $\bf B$.  We observe next
that Equations (\ref{1})-(\ref{5}) separate into two groups, namely
(\ref{1}), (\ref{3}), (\ref{5}) and (\ref{2}), (\ref{4}), involving only
components of $G$ which are odd and even in the momenta respectively.
This fact implies that the first integral can be written as $G = G_{odd} +
G_{even}$, where
 \begin{equation} \label{oe} \begin{array}{rcl} G_{odd}({\bf q}, {\bf p})
&=& L^{ijk}({\bf q})p_ip_jp_k + B^{k}({\bf q})p_{k},\\ [0.3cm]  G_{even}({\bf q},
{\bf p})& = & K^{ij}({\bf q})p_ip_j + U({\bf q}) \end{array}
 \end{equation}
 and furthermore that $\{H, G_{odd}\} = \{H,G_{even}\} =0.$
 \begin{defi} \label{d1} We say that a Hamiltonian system defined by
(\ref{H}) is {\em trivial} if it admits a  first
integral $G = X^ip_i$ linear in the momenta, where $\bf X$ is a Killing vector, satisfying
${\bf X}(V) = 0.$ Accordingly, we say that a Killing tensor $\bf L$ of
valence three is {\em trivial} if it is a product of the metric $\bf g$
and a Killing vector $\bf X$, i.e. ${\bf L} = {\bf g}\odot{\bf X}$.
 \end{defi}
These arguments confirm that (\ref{F}) is the most general
form of a first integral cubic in the momenta and a Hamiltonian system (\ref{H}) admitting a first integral of the
form (\ref{G}) is in fact {\it superintegrable}, provided it is not trivial, for in this case one can
 combine the metric $\bf g$ with $\bf X$ and square of $\bf X$ to obtain cubic and quadratic terms in the momenta respectively
 in the representation
(\ref{G}).  Note the decomposition (\ref{oe}) was proven in \cite{NY}
by employing a time reflection symmetry assumption that does not seem to be
required.

We note that this observation can be easily extended to the case of Hamiltonian systems of the
type (\ref{H}) admitting first integrals of order $r > 3$ in the momenta of the form:
\begin{eqnarray}
\label{genF}
\lefteqn{F_r ({\bf q}, {\bf p}) =} \\ [0.5cm] \nonumber
& & K_r^{a_1\ldots a_r}({\bf q})p_{a_1\ldots a_r} + \cdots + K_i^{a_1\ldots a_i}({\bf q})p_{a_1\ldots a_i} + \cdots + K_1({\bf q})^{a_1}p_{a_1} + U({\bf q}),
\end{eqnarray}
where $ 1 \le a_1, \ldots, a_r \le n,$ $ r>3$. Indeed, the vanishing of the poisson bracket $\{F_r, H\} = 0$ yields
in this case the following  tensor equations:
\begin{equation}
\begin{array}{c}
[{\bf K}_r, {\bf g}] = 0,\\[0.3cm]
[{\bf K}_{r-1}, {\bf g}] = 0, \\[0.3cm]
[{\bf K}_{r-2}, {\bf g}] = r {\bf K}_rdV, \\[0.3cm]
\vdots \\[0.5cm]
[{\bf K}_i, {\bf g}] = (i+2){\bf K}_{i+2}dV, \\ [0.3cm]
\vdots \\[0.5cm]
[{\bf K}_1, {\bf g}] = 3 {\bf K}_3 dV, \\ [0.3cm]
dU{\bf g} = [U,{\bf g}] = 2{\bf K}_2dV,\\[0.3cm]
{\bf K}_1(V) = 0,
\end{array}
\end{equation}
where the tensorial quantities ${\bf K}_r$, $r \ge 1$ are determined by the corresponding components of
(\ref{genF}). It is clear therefore that a general first integral $F_r$ of order $r$ given by (\ref{genF})
can also be written as $F_r = F_{r (odd)} + F_{r (even)}$, immediately implying
$\{H, F_{r (odd)}\} = \{H, F_{r (even)}\} = 0$. Accoordingly, we arrive at the following result:
\begin{prop}
A Hamiltonian  system with two degrees of freedom determined by (\ref{H}) which admits
a general first integral (\ref{genF}) of order $r \ge 3$ having both even and odd terms
in the momenta is necesserily superintegrable.
\end{prop}
\begin{coro}
Given a general first integral of order $r \ge 3$ of the type  (\ref{genF})
for a Hamiltonian system (\ref{H}). Then only the tensorial quantities
${\bf K}_r$ and ${\bf K}_{r-1}$ that define the first two terms of (\ref{genF})
are Killing tensors: ${\bf K}_r \in {\cal K}^r(M), {\bf K}_{r-1} \in {\cal K}^{r-1}(M)$.
\end{coro}
We conclude therefore
that for a Hamiltonian system (\ref{H}) admitting a cubic first integral (\ref{F})
 determined by its Killing tensor $\bf L$
and vector field $\bf B$, the following tensorial equations are essential:
 \begin{eqnarray} [{\bf L},{\bf g}] = 0, & \mbox{(Killing tensor equation
for $\bf L$)} \label{LE}\\ [0.3cm] [{\bf B},{\bf g}] = 3{\bf L}\mbox{d}V,
& \mbox{(recursion tensor equation for $\bf B$)} \label{BE}\\[0.3cm]
 {\bf B}(V) = 0.  & \mbox{(compatibility condition for $V$)} \label{VE}
 \end{eqnarray}
 Let us fix $\bf L$, $\bf B$ and $V$ in (\ref{F}) and (\ref{H})
respectively.  Since $(M, {\bf g})$ is of constant curvature, the
dimension of the corresponding isometry group $I(M)$ is three.  Therefore
the space ${\mathcal K}^1(M)$ of Killing vectors (infinitesimal isometries) of
$M$ is spanned by three Killing vectors ${\bf X}_1$, ${\bf X}_2$ and ${\bf
X}_3$.  Clearly, in view of linearity of the Schouten bracket, the vector
field $\tilde{\bf B}$ given by \begin{equation} \tilde{\bf B} = {\bf B} +
\sum_{i=1}^3k_i{\bf X}_i, \label{B1} \end{equation} is also a solution to
the tensor equation (\ref{BE}), where ${k}_1, {k}_2, {k}_3 \in {\mathbb
R}$ are arbitrary constants. Therefore as far as the equation (\ref{BE})
is concerned, each $\bf B$ generates a space $S_{\bf B} \subset TM$ of
vector fields satisfying (\ref{BE}), $S_{\bf B} = \{{\bf B}\}\cup {\mathcal
K}^1(M)$.  This is not true however for the equation (\ref{VE}).

We also note that the right hand side of (\ref{BE}) measures the deviation of the
vector field $\bf B$ from a Killing vector.

\begin{rem} {\rm The space of vector fields $S_{\bf B}$ is not closed
under the Lie commutator.  Indeed, let ${\bf B}_1, {\bf B}_2 \in S_{\bf
B}$. Hence, in view of (\ref{B1}) ${\bf B}_1 = {\bf B} + {\bf Y}_1$ and
${\bf B}_2 = {\bf B} + {\bf Y}_2,$ where ${\bf Y}_1, {\bf Y}_2 \in {\mathcal
K}^1(M)$. Taking into account (\ref{BE}) and (\ref{VE}) and using the
standard properties of Lie derivatives, we get $[[{\bf B}_1, {\bf
B}_2],{\bf g}] = 3 \tilde{\bf L}\mbox{d}V + 3{\bf L}\mbox{d}\tilde{V},$
where $\tilde{\bf L} = {\mathcal L}_{{\bf Y}_1-{\bf Y}_2}{\bf L}$ and
$\tilde{V} = ({\bf Y}_1-{\bf Y}_2)(V).$} \label{r1} \end{rem}

 \begin{rem} {\rm Lie derivative deformations of $\bf B$ with respect to
Killing vectors do not preserve the form of $\bf B$.  Indeed, using the
same arguments as in Remark \ref{r1}, it is easy to see, that $[[{\bf B},
{\bf X}],{\bf g}] = 3 \tilde{\bf L}\mbox{d}V + 3{\bf L}\mbox{d}\tilde{V},$
where ${\bf X} \in {\mathcal K}^1(M)$ and $\tilde{\bf L} = {\mathcal L}_{\bf
X}{\bf L}$, $\tilde{V} = {\bf X}(V).$ Note the right hand side of the last expession
is different from that of (\ref{BE})}.\end{rem}

 \begin{rem} \label{r3}{\rm Let ${\bf B} \in {\mathcal K}^1(M).$ This
immediately entails that the corresponding Hamiltonian system is trivial
and superintegrable, enjoying a second additional first integral of the
form $G = L^{ijk}p_ip_jp_j.$ The right hand side of (\ref{BE}) in this case is also
changed. } \end{rem}

 In what follows, we find and use isometry group invariants of Killing
tensors of valence three defined in the Euclidean plane ${\mathbb R}^2$ to
classify cubic integrals of motion (\ref{F}) up to the leading term represented by a Killing
tensor of valence three for
the Hamiltonian systems with two degrees of freedom defined in ${\mathbb
R}^2$.

 \section{Isometry group invariants of Killing tensors of valence three}

Let $(M, {\bf g})$ be the Euclidean plane ${\mathbb R}^2$.  In this
section we apply the method developed recently by the authors in \cite{MST, MST1, MST2, MST3} to the problem of classification of Killing tensors belonging to
the linear space ${\mathcal K}^3({\mathbb R}^2)$.

Consider the linear space ${\mathcal K}^3({\mathbb R}^2)$. The dimension of the space can be
computed by employing the Delong-Takeuchi-Thompson formula \cite{De,Ta,Th}:
$$\mbox{dim}{\mathcal K}^3({\mathbb R}^2) = \frac{1}{2}
{2 + 3 \choose 3+1}{2 + 3 - 1 \choose 3} = 10. $$
 Alternatively, we can solve the Killing
tensor equation $[{\bf L}, {\bf g}] = 0, $ ${\bf L} \in  {\mathcal K}^3({\mathbb R}^2)$ for
the metric tensor of ${\mathbb R}^2$ in (say) Cartesian coordinates $(x,y)$, to get
\begin{equation}
\label{KTE}
\begin{array}
{rcl}
 L^{111} &=& a_1 + 3 \alpha_1 y + 3 \beta_1 y^2 + \gamma y^3,\\ [0.3cm]
 L^{112} &=& a_2 + \alpha_2 y - \alpha_1 x - 2 \beta_1 xy +
\beta_2 y^2 - \gamma xy^2,\\ [0.3cm]
 L^{122} &=& a_3 - \alpha_2 x - \alpha_3 y - 2 \beta_2 xy +
\beta_1 x^2 + \gamma yx^2,\\[0.3cm]
 L^{222} &=& a_4 + 3 \alpha_3 x + 3 \beta_2 x^2 - \gamma x^3.
 \end{array}
 \end{equation}
 The ten parameters $a_1, \ldots, \gamma$ are the constants of integration
which  also imply that the dimension of ${\mathcal K}^3({\mathbb R}^2)=10$.
Accordingly, we can view each element of ${\mathcal K}^3({\mathbb R}^2)$ as an
algebraic object determined by its ten parameters $a_1, \ldots, \gamma$.
Furthermore, for the linear space ${\mathcal K}^3({\mathbb R}^2)$, we can
formulate an analogue of the main problem of the classical theory of
algebraic invariants (see \cite{Hil,MST3} for more details). Indeed, recall that
the main problem of the classical theory of algebraic invariants developed by
Hilbert \cite{Hil} in  modern mathematical language reads:
\begin{prob}
 Determine the linear action of a group $G$ on a $K$-vector space $V$. Then in the ring of polynomial
functions $K[V]$ describe the subring $K[V]^G$ of all polynomial functions on $V$ which are invariant under
the action of $G$ (i.e., the {\em invariants}).
\end{prob}
In the present study, we formulate the problem as follows:
 \begin{prob} Let $\Sigma$ be the space spanned by the ten parameters $a_1,
\ldots, \gamma$ that appear in (\ref{KTE}). Determine the action induced
by the isometry group $I({\mathbb R}^2)$ on $\Sigma$. Then in the space of
functions defined in $\Sigma$ describe the subspace of all functions in
$\Sigma$ which are invariant under the action induced by the isometry
group.  \label{p1}
 \end{prob}
 Note that the linear spaces $\Sigma$ and ${\mathcal K}^3({\mathbb R}^2)$ are
isomorphic. The solution of Problem $\ref{p1}$ consists of two essential parts:
 first determining the action of $I({\mathbb R}^2)$ in $\Sigma$ and
second  finding the invariants. Unlike  the classical theory of algebraic
invariants \cite{Hil}, where the group acts naturally in a space of
polynomials by linear substitutions,  the  determination of
the action of the isometry group in the present situation is more complex.  To make the first step,
we proceed as in the analogous problems considered by the authors in
\cite{MST, MST2, MST3} for the linear spaces ${\mathcal K}^2({\mathbb R}^2)$
and ${\mathcal K}^2({\mathbb R}_1^2)$, where ${\mathbb R}_1^2$ denotes the
Minkowski plane. Observe that the generators of the Lie algebra
$i({\mathbb R}^2)$ of the isometry group $I({\mathbb R}^2)$ with respect
to Cartesian coordinates $(x,y)$ are the vector fields ${\bf X} =
\frac{\partial}{\partial x},$ ${\bf Y} = \frac{\partial}{\partial y}$ and
${\bf R} = y\frac{\partial}{\partial x} - x\frac{\partial}{\partial y}$
whose flows represent translations along $x$, $y$ and a rotation
respectively. They enjoy the following commutator relations:
 \begin{equation} \label{XYR} [{\bf X}, {\bf Y}] = 0, \quad [{\bf X}, {\bf
R}] = - {\bf Y}, \quad [{\bf Y}, {\bf R}] = {\bf X}.
 \end{equation}
 Now we introduce the {\it projection map} $\pi: {\mathcal K}^3({\mathbb R}^2)
\rightarrow T\Sigma$ defined for a fixed ${\bf L}^0 \in {\mathcal
K}^3({\mathbb R}^2)$ as follows:
 \begin{equation}
\label{pm}
\pi({\bf L}^0) = \sum_{i=1}^4a_i^0\frac{\partial}{\partial a_i} + \sum_{i=1}^3
\alpha_i^0\frac{\partial}{\partial \alpha_i} + \sum_{i=1}^2\beta_i^0\frac{\partial}{\partial \beta_i} +
 \gamma^0\frac{\partial }{\partial \gamma},
 \end{equation}
where $a_1^0, \ldots, \gamma^0$ are the ten parameters that determine ${\bf L}^0$ via the representation
(\ref{KTE}) with respect to Cartesian coordinates $(x,y)$ and $a_1, \ldots \gamma$ are the ten coordinate
functions of the space $\Sigma$. In order to determine the induced action of $I({\mathbb R}^2)$
in $\Sigma$, we use the composition map $\pi \circ {\mathcal L}$ of $\pi$ and the Lie derivative operator
${\mathcal L}$. Define the following vector fields in $T\Sigma$:
\begin{equation}
\label{VS}
{\bf V}_1 := \pi{\mathcal L}_{\bf X}{\bf L}, \quad {\bf V}_2 := \pi{\mathcal L}_{\bf Y}{\bf L}, \quad {\bf V}_3 := \pi{\mathcal L}_{\bf R}{\bf L},
\end{equation}
 where $\bf L$ is the general Killing tensor, whose components with
respect to Cartesian coordinates $(x,y)$ are given in (\ref{KTE}). Note
that the Lie derivative deformations of $\bf L$ with respect to Killing
vectors are themselves Killing tensors of valence three and as such
represent elements of the linear space ${\mathcal K}^3({\mathbb R}^2)$.  This
fact is a consequence of the Jacobi identity for the Schouten bracket
acting in the space of symmetric contravariant tensorial quantities. We
conclude therefore that the formulas (\ref{VS}) are well-defined.
Employing the formulas (\ref{VS}) in conjunction with (\ref{KTE}), we
arrive at the following representations of ${\bf V}_1,$ ${\bf V}_2$, ${\bf
V}_3$ $\in T\Sigma$ with respect to the coordinate functions $a_1, \ldots,
\gamma$:
 \begin{equation}
\label{VS1}
\begin{array}{rcl}
    {\bf V}_1 &=& \displaystyle -\alpha_1 \frac{\partial}{\partial a_2} - \alpha_2 \frac{\partial}{\partial a_3} + 3\alpha_3 \frac{\partial}{\partial a_4} -2\beta_1 \frac{\partial}{\partial \alpha_2} + 2\beta_2 \frac{\partial}{\partial \alpha_3} - \gamma \frac{\partial}{\partial \beta_2}\\ [0.3cm]
    {\bf V}_2 &=& \displaystyle 3\alpha_1 \frac{\partial}{\partial a_1} + \alpha_2 \frac{\partial}{\partial a_2} -\alpha_3 \frac{\partial}{\partial a_3} + 2\beta_1 \frac{\partial}{\partial \alpha_1} + 2\beta_2 \frac{\partial}{\partial \alpha_2} + \gamma \frac{\partial}{\partial \beta_1}\\ [0.3cm]
    {\bf V}_3 &=& \displaystyle 3 a_2\frac{\partial}{\partial a_1} + (2 a_3-a_1) \frac{\partial}{\partial a_2} + (a_4 - 2 a_2) \frac{\partial}{\partial a_3} - 3a_3 \frac{\partial}{\partial a_4} \\ [0.3cm]  & & \displaystyle + \alpha_2 \frac{\partial}{\partial \alpha_1} - 2(\alpha_1 + \alpha_3) \frac{\partial}{\partial \alpha_2} + \alpha_2 \frac{\partial}{\partial \alpha_3} + \beta_2 \frac{\partial}{\partial \beta_1} - \beta_1 \frac{\partial}{\partial \beta_2}
 \end{array}
 \end{equation}
Our next observation is that ${\bf V}_1, {\bf V}_2, {\bf V}_3$ satisfy the following commutator relations:
\begin{equation}
\label{CR}
 [{\bf V}_1, {\bf V}_2] = 0, \quad [{\bf V}_1, {\bf V}_3] = - {\bf V}_2, \quad
[{\bf V}_2, {\bf V}_3] = {\bf V}_1.
\end{equation}
Therefore ${\bf V}_1, {\bf V}_2, {\bf V}_3$  form a basis for a Lie algebra $i_{{\mathcal K}^3({\mathbb R}^2)}(\Sigma)$, which
is a Lie subalgebra of $T\Sigma$. Furthermore, its generators ${\bf V}_1, {\bf V}_2$ and ${\bf V}_3$ satisfy
the same commutator relations as the generators ${\bf X}, {\bf Y}$ and ${\bf R}$ of $i({\mathbb R}^2) =
{\mathcal K}^1({\mathbb R}^2)$ in $T{\mathbb R}^2$ (compare (\ref{XYR}) with (\ref{VS1})). Therefore we have
proven the following
\begin{prop}
\label{prop1}
The vector space $i_{{\mathcal K}^3({\mathbb R}^2)}(\Sigma) \subset T\Sigma$ is a Lie subalgebra of $T\Sigma$ isomorphic
to the Lie algebra of Killing vectors $i({\mathbb  R}^2) = {\mathcal K}^1({\mathbb R}^2)$.
\end{prop}
Proposition \ref{prop1} establishes the induced action of $I({\mathbb R}^2)$ in $\Sigma \simeq {\mathcal K}^3({\mathbb R}^2)$.
Therefore we have solved the first part of Problem \ref{p1}.

Recall that invariance of an object under entire Lie group is equivalent to infinitesimal invariance under
infinitesimal generators of the corresponding Lie algebra. This observation
and the result of Proposition \ref{prop1} prompts the following
 \begin{defi} \label{d2} A smooth function $F: \, \Sigma \rightarrow
{\mathbb R}$ is said to be an {\em $I({\mathbb R}^2)$-invariant of ${\mathcal
K}^3({\mathbb R}^2)$} iff
 \begin{equation}
{\bf V}_i (F) = 0, \quad i = 1, 2, 3,
\label{PDES}
 \end{equation}
 where ${\bf V}_i,$ $i = 1,2,3$ are the generators of the Lie algebra
$i_{{\mathcal K}^3({\mathbb R}^2)}(\Sigma)$ isomorphic to the Lie algebra
$i({\mathbb R}^2) = {\mathcal K}^1({\mathbb R}^2)$ of Killing vectors defined
on
${\mathbb R}^2$.
 \end{defi}
 Solving the second part of Problem \ref{p1} comes down to solving the
system of partial differential equations (\ref{PDES}) determined by the generators of $i_{{\mathcal
K}^3({\mathbb R}^2)}(\Sigma)$. A simple counting argument reveals that
there are 7 = 10 (dimension of $\Sigma$) - 3 (dimension of the Lie group
$I({\mathbb R}^2)$) essential invariants that determine the space of all
$I({\mathbb R}^2)$-invariants of ${\mathcal K}^3({\mathbb R}^2)$. In what
follows we may not need to determine all of them, since we wish to consider Problem
\ref{p1} in conjunction with the dynamical problem envisaged in the
Introduction.  More specifically, we will exclude from further
consideration all Killing tensors that are
trivial, according to Definition \ref{d1}.  Ultimately, this will
lead to a different (smaller) set of essential $I({\mathbb
R}^2)$-invariants of the linear space ${\mathcal K}^3({\mathbb R}^2)$.

\section{Essential integral submanifolds of the space  $\Sigma$}

In this section we carry over the ideas exhibited in the previous section
to the classification problem of the linear space ${\mathcal K}^3({\mathbb
R}^2)$, that is find and use $I({\mathbb R}^2)$-invariants of ${\mathcal
K}^3({\mathbb R}^2)$ to classify Killing tensors of valence three and
thus, the corresponding  first integrals of motions of
(\ref{H}) up to their leading terms.

First, we wish to exclude from further consideration the Killing tensors
corresponding to the trivial  first integrals (see
Definition \ref{d1}). In Cartesian coordinates the most general Killing
tensor ${\bf L}_{tr} \in {\mathcal K}^3({\mathbb R}^2)$ of this type can be
obtained via multiplication of the metric $\bf g$ of ${\mathbb R}^2$ by
the most general Killing vector ${\bf X}_{gen}$ of ${\mathbb R}^2$ spanned
by $\bf X$, $\bf Y$ and $\bf R$: $${\bf X}_{gen} = (a +
cy)\frac{\partial}{\partial x} + (b -cx)\frac{\partial}{\partial y}, \quad
a,b,c \in {\mathbb R}.$$ Thus, the components of ${\bf L}_{tr}$ are given
by $L^{ijk}_{tr} = g^{(ij}X^{k)}_{gen}$. Comparing the last formula with
(\ref{KTE}) and taking into account $g^{12} = 0, g^{11} = g^{22} = 1$, we
come to the conclusion that in $\Sigma$ the subspace ${\mathcal T} \subset
\Sigma$ corresponding to the space of trivial Killing tensors in ${\mathcal
K}^3({\mathbb R}^2)$ is determined by the following conditions:
 \begin{equation}
    a_1 - 3 a_3 = 0, 3 a_2 - a_4 = 0, \alpha_1 + \alpha_3 = 0, \alpha_2=0,
\beta_1= \beta_2 =0, \gamma=0.
    \label{conds}
 \end{equation}
 It is easy to see that the space ${\mathcal T} \subset \Sigma$ defined by
(\ref{conds}) is an integral submanifold in $\Sigma$ whose tangent space
$T{\mathcal T}$ is spanned by the vector fields
 \begin{equation}
\label{TTT}
{\bf T}_1  = 3 \frac{\partial}{\partial a_1}  + \frac{\partial}{\partial a_3},
\quad
 {\bf T}_2 = \frac{\partial}{\partial a_2} + 3\frac{\partial}{\partial a_4},
 \quad
  {\bf T}_3 = \frac{\partial}{\partial \alpha_1} - \frac{\partial}{\partial \alpha_3}.
\end{equation}
Indeed, the system $\{{\bf T}_1, {\bf T}_2, {\bf T}_3\}$
is in involution ($\{{\bf T}_1, {\bf T}_2, {\bf T}_3\}$ is a basis
of an abelian Lie algebra in $T\Sigma$), and as such, in view of
Frobenius' theorem, is integrable. Interestingly, the system
\begin{equation}
\label{SVT}
\{{\bf V}_1, {\bf V}_2, {\bf V}_3, {\bf T}_1, {\bf T}_2, {\bf T}_3\}
\end{equation}
is a basis of a six-dimensional Lie algebra in $T\Sigma$ with the following
commutator table:
\[
 \begin{array}{c||c|c|c|c|c|c}
    & {\bf V}_1 & {\bf V}_2 & {\bf V}_3 & {\bf T}_1 & {\bf T}_2 & {\bf T}_3\\ \hline \hline
  {\bf V}_1 & 0 & 0 & -{\bf V}_2 & 0 & 0 & {\bf T}_2\\ \hline
  {\bf V}_2 & 0 & 0 & {\bf V}_1 & 0 & 0 & -{\bf T}_1\\ \hline
  {\bf V}_3 & {\bf V}_2 & -{\bf V}_1 & 0 & {\bf T}_2 & -{\bf T}_1 & 0\\ \hline
  {\bf T}_1 & 0 & 0 & -{\bf T}_2 & 0 & 0 & 0\\ \hline
  {\bf T}_2 & 0 & 0 & {\bf T}_1 & 0 & 0 & 0\\ \hline
  {\bf T}_3 & -{\bf T}_2 & {\bf T}_1 & 0 & 0 & 0 & 0
  \end{array}
 \]
Therefore the system of vector fields  (\ref{SVT}) is also integrable.
According to Definition \ref{d2}, in order to find essential $I({\mathbb R}^2)$-invariants
of ${\mathcal K}^3({\mathbb R}^2)$ of {\em non-trivial} Killing tensors, we
have to solve the following system of PDE's:
\begin{equation}
{\bf T}_i (F) = 0, \quad {\bf V}_i(F) = 0, \quad i = 1,2,3
\label{TV}
\end{equation}
for a smooth function $F: \, \Sigma \rightarrow {\mathbb R}$. The equations
$F = c,$ where $c$ is a constant, will determine integral submanifolds of the integrable
distribution of vector fields (\ref{SVT}).

Although in principle the system (\ref{TV})  can be solved by the method of characteristics, the actual
 execution of the
procedure is  an arduous computational task. To alleviate the difficulties, we solve (\ref{TV})
in two steps. More specifically, let us first find all essential $I({\mathbb R}^2)$-invariants
of ${\mathcal K}^3({\mathbb R}^2)$ for a subsystem of PDE's, namely the following.
\begin{equation}
\label{TV1}
{\bf T}_i (F) = 0, \quad {\bf V}_j(F) = 0, \quad i = 1,2,3 \quad j = 1, 2.
\end{equation}
Employing the method of characteristics, we find the following set of five essential
$I({\mathbb R}^2)$-invariants of ${\mathcal K}^3({\mathbb R}^2)$ for the system (\ref{TV1})
as functions of the ten parameters $a_1, \ldots, \gamma$:
 \begin{equation}
 \begin{array}{rcl}
    \tilde{I}_1 &=& \gamma \alpha_2 - 2 \beta_1 \beta_2,\\ [0.3cm]
    \tilde{I}_2 &=& \gamma(\alpha_1 + \alpha_3) + \beta_2^2 - \beta_1^2,\\ [0.3cm]
    \tilde{I}_3 &=& \gamma^2 (a_1 - 3 a_3) + 3 \gamma (\beta_2 \alpha_2 - \beta_1 (\alpha_1 + \alpha_3))  + 2 \beta_1 (\beta_1^2 - 3 \beta_2^2),\\ [0.3cm]
    \tilde{I}_4 &=& \gamma^2 (3a_2 - a_4) - 3 \gamma (\beta_1 \alpha_2 + \beta_2 (\alpha_1 + \alpha_3)) + 2 \beta_2 (3\beta_1^2 - \beta_2^2),\\ [0.3cm]
    \tilde{I}_5 &=& \gamma,
 \end{array}
 \label{EI}
 \end{equation}
Note that all of the invariants (\ref{EI}) are {\em homogeneous polynomials}. Therefore
we conclude that the most general $I({\mathbb R}^2)$-invariant $\tilde{F}: \Sigma \rightarrow {\mathbb R}$
satisfying (\ref{TV1}) is given by
\begin{equation}
\tilde{F} = \tilde{F}(\tilde{I}_1, \tilde{I}_2,\tilde{I}_3,\tilde{I}_4,\tilde{I}_5).
\end{equation}
 Now let us treat the functions $\tilde{I}_i, i = 1, \ldots, 5$ as new
coordinates and apply the last vector field ${\bf V}_3$ to $\tilde{F}$.
This leads to the first order linear partial differential equation
 \begin{equation}
0 = -2\tilde{I}_2\frac{\partial \tilde{F}}{\partial \tilde{I}_1} + 2\tilde{I}_1\frac{\partial \tilde{F}}{\partial \tilde{I}_2}
+ 3\tilde{I}_4\frac{\partial \tilde{F}}{\partial \tilde{I}_3} - 3\tilde{I}_3\frac{\partial \tilde{F}}{\partial \tilde{I}_4}
 \end{equation}
and four essential $I({\mathbb R}^2)$-invariants of ${\mathcal K}^3({\mathbb R}^2)$ as functions
of $\tilde{I}_i, i=1,\ldots, 5$ that are found to be:
 \begin{equation}
\label{AI}
\begin{array}{rcl}
I_1 & = & 2(\tilde{I}_1^2-3\tilde{I}_2^2)\tilde{I}_1\tilde{I}_3\tilde{I}_4
+(3\tilde{I}_1^2 - \tilde{I}_2^2)\tilde{I}_2(\tilde{I}_3^2 - \tilde{I}_4^2), \\ [0.3cm]
I_2 & = & \tilde{I}_1^2 + \tilde{I}^2_2,\\ [0.3cm]
I_3 & = & \tilde{I}_3^2 + \tilde{I}_4^2, \\[0.3cm]
I_4 & =& \tilde{I}_5.
\end{array}
\end{equation}
 This result puts in evidence that the function
 \begin{equation}\tilde{F} =
\tilde{F}({I}_1, {I}_2, {I}_3, {I}_4)\end{equation}
is the solution to Problem
\ref{p1} for the {\em non-trivial} elements of ${\mathcal K}^3({\mathbb
R}^2)$, that is the Killing tensors that determine non-trivial, according
to Definition \ref{d1},  first integrals of (\ref{H}) cubic in the momenta.
Now we can use the {\em essential $I({\mathbb R}^2)$-invariants of ${\mathcal
K}^3({\mathbb R}^2)$} ${I}_i, i = 1,\ldots, 4$ (\ref{AI}) to classify
non-trivial Killing tensors of valence three up to the action of the
isometry group $I({\mathbb R}^2)$ according to whether ${I}_i, i
=1,\ldots, 4$ are equal to zero or not. Therefore there are $2^4 = 16$
distinct classes of non-trivial Killing tensors characterized by ${I}_i, i
=1,\ldots, 4$ (\ref{AI}). The elements of each class are equivalent up to
the action of the isometry group $I({\mathbb R}^2)$.  We note however that
the Killing tensors of valence three of the first integrals isolated
by Holt \cite{Ho} and Fokas-Lagerstr\"{o}m \cite{FL} are {\em
non-trivial}, yet the essential $I({\mathbb R}^2)$-invariants ${I}_i, i =
1,\ldots, 4$ all vanish identically (see the next section).  This happens
due to the fact that for the Killing tensors of both Holt and
Fokas-Lagerstr\"{o}m  first integrals, cubic in the momenta the
parameters $\gamma$, $\beta_1$ and $\beta_2$ vanish identically, which, in
turn, kills all of the essential $I({\mathbb R}^2)$-invariants ${I}_i, i =
1,\ldots, 4$ given by (\ref{AI}).  As might be expected, we need extra
invariant(s) to distinguish the cases when $\gamma = \beta_1 = \beta_2
=0$. Indeed, observe that $\gamma$ is an essential invariant of ${\mathcal
K}^3({\mathbb R}^2)$ for non-trivial Killing tensors of valence three.
Therefore we can consider the integral hypersurface of the space $\Sigma
\backslash {\mathcal T}$ determined by the condition $\gamma = 0$. Using the
same arguments as above, we find that the homogeneous polynomial
$\beta_1^2 + \beta^2_2$ is an $I({\mathbb R}^2)$-invariant of ${\mathcal
K}^3({\mathbb R}^2)$ in this case. Therefore we can consider the integral
submanifold $\Sigma_1\backslash {\mathcal T}$ of $\Sigma \backslash {\mathcal T}$
determined by the conditions: $\gamma = 0,$ $ \beta_1^2 + \beta^2_2 = 0$.
It is easy to see by using the same counting argument as that given in the end
of Section 3 that there exists
only {\em one} essential $I({\mathbb R}^2)$-invariant of ${\mathcal
K}^3({\mathbb R}^2)$ in the subspace $\Sigma_1\backslash {\mathcal T}$.
Employing the procedure described above, we find it to be the following
homogeneous polynomial of the parameters $\alpha_1, \alpha_2$ and
$\alpha_3$:
 \begin{equation}
 I_* = (\alpha_1 + \alpha_3)^2 + \alpha_2^2.
 \label{*}
 \end{equation}
 Now we can classify Killing tensors of valence three in the space
$\Sigma_1\backslash {\mathcal T}$ up to the action of the isometry group
$I({\mathbb R}^2$) according to whether $I_*$ zero or not. In view of
homogeneity of the polynomial $I_*$ the Killing tensors of valence three
determined by the condition $I_* \not= 0 $ are equivalent up to rescaling
$\tilde{x} = kx, \tilde{y} = ky,$ $k \in {\mathbb R}$ (see (\ref{KTE}))
that preserves the signature of the metric $\bf g$. Note also that there
exist no other essential $I({\mathbb R}^2)$-invariants of ${\mathcal K}^3({\mathbb R}^2)$
involving $\alpha_1, \alpha_2$ and $\alpha_3$. Thus, there are only two
equivalence classes of Killing tensors in $\Sigma_1\backslash {\mathcal T}$. Let
$\Sigma_2 = \Sigma \backslash \Sigma_1$.
We summarize our observations in the following table.

 \begin{table}[ht]
 \begin{center}
 \begin{tabular}{|c|c|c|c|}\hline
 {\small Case} & \begin{tabular}{c} {\small Integral } \\[0.1cm] {\small submanifold in} $\Sigma$
 \end{tabular} & {\small Invariants }& \begin{tabular}{c} $\#$ {\small of
equivalence} \\[0.1cm]
 {\small classes of elements} \\[0.1cm]
  of ${\mathcal K}^3({\mathbb R}^2)$ \end{tabular} \\[0.2cm] \hline
 \begin{tabular}{c} \\$\gamma^2 +\beta_1^2 + \beta_2^2 \not=0$ \\ [0.2cm] \end{tabular}  & $\Sigma_2\backslash {\mathcal T}$
& ${I}_1, {I}_2, {I}_3, {I}_4$ & $2^4-6 = 10$ \\[0.2cm]\hline
 \begin{tabular}{c} \\ $\gamma^2 +\beta_1^2 + \beta_2^2 =0$ \\[0.2cm]
 \end{tabular} & $\Sigma_1\backslash {\mathcal T}$ & $I_*$ & $2$ \\ [0.2cm]
\hline
 \end{tabular}
 \caption{\small Group invariant classification of Killing tensors of valence three
of non-trivial cubic in the momenta first integrals of motion}
 \label{Tab}
 \end{center}
 \end{table}
\begin{rem} {\rm Note  the conditions $I_2 = I_4 =0$ and $I_3=I_4=0$ incite
the condition $\gamma= \beta_1 =  \beta_2  = 0$. Therefore we have reduced the
number of equivalence classes corresponding to the condition $\gamma^2
+\beta_1^2 + \beta_2^2 \not=0$ by 6, which is the number of cases including the
conditions $I_2 = I_4 =0$ and/or $I_3=I_4=0$}
\end{rem}

 In the next section we apply our classification scheme to study the
cubic in the momenta first integrals of motion isolated by Holt \cite{Ho} and
Fokas-Lagerstr\"{o}m \cite{FL}.

\section{The Holt and Fokas-Lagerstr\"{o}m potentials}
Consider the Holt and Fokas-Lagerstr\"{o}m integrable systems characterized by the existence of the  first integrals of motion
 \begin{eqnarray}
\lefteqn{F_{H}({\bf q},{\bf p}) = } \label{Ho} \\[0.3cm]
& & 2p_1^3 + 3p_1p_2^2 + 3p_1(-3(q^2)^2 + 2(q^1)^2 +  2\delta)(q^2)^{-2/3} + \\ [0.3cm]
& &  18p_2q^1(q^2)^{1/3}, \nonumber
 \end{eqnarray}
and
 \begin{eqnarray}
\lefteqn{F_{FL} ({\bf q},{\bf p}) = } \label{FL} \\ [0.3cm]
&  & (p_1^2 - p_2^2)(q^1p_2 - q^2p_1) - \\[0.3cm] & &  4(q^2p_1 + q^1p_2)((q^1)^2 -(q^2)^2)^{-2/3}, \nonumber
 \end{eqnarray}
 respectively given in the standard position-momenta coordinates.
Comparing the cubic  terms in $F_H$ and $F_{FL}$ with the
general formulas (\ref{KTE}) and taking into account the conditions
(\ref{conds}), we conclude that the corresponding Killing tensors of
valence three ${\bf L}_{H}$ and ${\bf L}_{FL}$ determining the leading
terms of (\ref{Ho}) and (\ref{FL}) respectively are non-trivial, yet they
both satisfy the invariant condition $\gamma = \beta_1 = \beta_2 = 0$.
Therefore both ${\bf L}_{H}$ and ${\bf L}_{FL}$ belong to the space
$\Sigma_1\backslash {\mathcal T}$ and as such can be characterized by the
essential $I({\mathbb R}^2)$-invariant $I_*$ given by (\ref{*}). Indeed,
comparing (\ref{Ho}) and (\ref{FL}) with (\ref{KTE}), we get for ${\bf
L}_H$:  $I_* = 0$ and for ${\bf L}_{FL}$: $I_*= 4/9$.  Hence, each of
${\bf L}_H$ and ${\bf L}_{FL}$ span the two equivalence classes of
non-trivial Killing tensors of valence three in $\Sigma_1\backslash {\mathcal
T}$. In other words, any Killing tensor belonging to $\Sigma_1\backslash
{\mathcal T}$ can be obtained from either ${\bf L}_{H}$ or ${\bf L}_{HL}$ via
the translations and rotations of the coordinates $(q^1,q^2)$ determined
by the generators $\bf X$, $\bf Y$ and $\bf R$ respectively or by rescaling.

Therefore any new integrable cases of (\ref{H}) admitting  first integrals of motion that are cubic in the momenta
within the space $\Sigma_1\backslash {\mathcal T}$ can only be found by varying the respective
vectors fields ${\bf B}_{H}$ and ${\bf B}_{FL}$ in (\ref{Ho}) and (\ref{FL})  and finding the corresponding
potential functions $V$ of (\ref{H}) via the formulas (\ref{BE}) and (\ref{VE}).
\section{Concluding remarks}
In this paper we have extended the isometry group invariants approach
developed in \cite{MST, MST1, MST2, MST3}  to the study of
Killing tensors of valence three in the Euclidean plane, or elements of
the linear space ${\mathcal K}^3({\mathbb R}^2).$ More specifically, Problem
\ref{p1}, which is an analogue of the main problem of the classical theory
of algebraic invariants \cite{Hil}, has been solved for the linear space of
non-trivial Killing tensors of valence three defined in the Euclidean
plane.

The approach allows us to classify  first integrals cubic in the momenta for
Hamiltonian systems defined in the Euclidean plane up to their leading
cubic terms by employing essential $I({\mathbb R}^2)$-invariants of ${\mathcal
K}^3({\mathbb R}^2)$.  We have studied the Killing tensors ${\bf
L}_{H}$ and ${\bf L}_{FL}$ of valence three determined by the  first
integrals of motion found by Holt and Fokas-Lagerstr\"{o}m respectively
and described the integral submanifold in the space $\Sigma$ determined by
the parameters of the generic Killing tensor of valence three defined in
the Euclidean plane, that contains both of them. It has also been
demonstrated that all other Killing tensors in this space are related to
${\bf L}_{H}$ and ${\bf L}_{FL}$ via the rigid motions of the Euclidean
space ${\mathbb R}^2$.

We are convinced that the method used in this paper to study Hamiltonian
systems can be successfully employed in other areas of mathematical
physics. Recall, that Delong \cite{De} studied the ten-parameter conformal
symmetry group of the two-dimensional wave equation. He has proved that
there are $(2k+1)(2k+2)(2k+3)/6$ independent $k$-order symmetries
generated by the ten basic elements of this conformal symmetry group. This
formula is an analogue of the Delong-Takeuchi-Thompson formula \cite{De, Ta,Th} used
in the present study.  Thus, one can derive invariants of the
ten-parameter conformal symmetry group in the space, say, of second-order
symmetries generated by the basic elements and use them in
classification-type problems that arise in the study of conformal
symmetries of the two-dimensional wave equation.

\bigskip

\noindent {\bf Acknowledgements.}  The authors are grateful to  Sergio Benenti,
Claudia Chanu and Bennno Fuchssteiner for illuminating discussions and useful
remarks. We also wish to thank Willard Miller for bringing to their attention
the reference
\cite{De}. The research was supported in part by the National Sciences and
Engineering Research Council of Canada (RGM, DT) and Alexander von Humboldt
Foundation (RGS).


\begin{thebibliography}{99}

\bibitem{Dr} J.  Drach, ``Sur l'int\'{e}gration logique des \'{e}quations de la dynamique \`{a} deux variables.  Forces
conservatives.  Int\'{e}grales cubiques.  Mouvement dans le plan,'' Compt.  Rend.  {\bf 200}, 22--26 (1935).


\bibitem{FL} A.  S.  Fokas and P.  A.  Lagerstr\"{o}m, ``Quadratic and cubic invariants in Classical Mechanics,'' J.  Math.
Anal.  Appl.  {\bf 74}, 325--341 (1980).



\bibitem{Ho} C.  R.  Holt, ``Construction of new integrable Hamiltonians in two degrees of freedom,'' J.  Math.  Phys.  {\bf
23}, 1037--1046 (1982).


\bibitem{NY} K.  Nakagawa and H.  Yoshida, ``A list of all integrable two-dimensional homogeneous polynomial potentials with a
polynomial integral of order at most four in the momenta,'' J.  Phys.  A:  Math.  Gen.  {\bf 34}, 8611--8630 (2001).

 \bibitem{FMSUW} J.  Fri\v{s}, V.  Mandrosov, Ya.  A.  Smorodinsky, M.  Uhl{i}{r} and P.  Winternitz, ``On higher symmetries
in quantum mechanics,'' Phys.  Lett.  {\bf 16}, 354--356 (1965).

\bibitem{FMSUW1} P. Winternitz, Ya. A. Smorodinsky, M.  Uhl{i}{r} and J.  Fri\v{s}, ``Symmetry groups in classical
and quantum mechanics,'' Soviet. J. Nuclear Physics {\bf 4}, 444--450 (1967).

\bibitem{Ra} M. F. Ra\~{n}ada, ``Superintegrable $n=2$ systems, quadratic constants of motion, and
potentials of Drach,'' J. Math. Phys. {\bf 38}, 4165--4178 (1997).


\bibitem{Ts} A.  V.  Tsiganov, ``The Drach superintegrable systems,'' J.  Phys.  A.  {\bf 33}, 7407--7422 (2000).

\bibitem{WG} S. Gravel and P. Winternitz, ``Superintegrability with third-order integrals in quantum and classical
mechanics,'' J. Math. Phys. {\bf 43}, 5902--5912 (2002).


 \bibitem{MST} R.  G.  McLenaghan, R.  G.  Smirnov and D.  The, ``Group invariant classification of separable Hamiltonian
systems in the Euclidean plane and the $O(4)$-symmetric Yang-Mills theories of Yatsun,'' J.  Math.  Phys.  {\bf 43},
1422--1440 (2002).

 \bibitem{MST1} R.  G.  McLenaghan, R.  G.  Smirnov and D.  The, ``The 1881 problem of Morera revisited,'' J.  Diff.  Geom.
Appl., in press.

 \bibitem{MST2} R.  G.  McLenaghan, R.  G.  Smirnov and D.  The, ``Group invariants of Killing tensors in the Minkowski
plane'', to appear in the Proceedings of ``Symmetry and Perturbation Theory - SPT2002'', the conference held in Cala Gonone,
19-26 May 2002, S.  Abenda, G.  Gaeta and S.  Walcher eds., World Scientific, 2003.

\bibitem{MST3} R.  G.  McLenaghan, R.  G.  Smirnov and D.  The, ``An extension of the classical theory of algebraic invariants
to pseudo-Riemannian geometry and Hamiltonian mechanics,''preprint, 63 pages, (2002).

\bibitem{Hil} D. Hilbert, {\it Theory of Algebraic Invariants}, (Cambridge University Press, 1993).

\bibitem{Ol} P. J. Olver, {\it Classical Invariant  Theory}, London Mathematical Society,
Student Texts {\bf 44}, (Cambridge University Press, 1999).

\bibitem{De} R. P. Delong, Jr. {\it Killing tensors and Hamilton-Jacobi equation}, PhD thesis, University of Minnesota, 1982.


 \bibitem{Ta} M. Takeuchi, ``Killing tensor fields on spaces of constant
curvature'', { Tsukuba J. Math.} {\bf 7}, 233--255 (1983).

 \bibitem{Th} G. Thompson, ``Killing tensors in spaces of constant
curvature'', { J. Math. Phys.} {\bf 27},  2693--2699 (1986).


  \end{thebibliography}
 \end{document}